# Super-resolution Reconstruction Algorithms Based on Fusion of Deep Learning Mechanism and Wavelet

Qi Zhang
16152010819@mail.ncut.edu.cn

Huafeng Wang*
wanghuafeng@ncut.edu.cn

Tao du, Sichen Yang,
Yuehai wang, Wenle Bai, Yang Yi

## ABSTRACT
In this paper, we consider the problem of super-resolution reconstruction. This is a hot topic because super-resolution reconstruction has a wide range of applications in the medical field, remote sensing monitoring, and criminal investigation. Compared with traditional algorithms, the current super-resolution reconstruction algorithm based on deep learning greatly improves the clarity of reconstructed pictures. Existing work like Super-Resolution Using a Generative Adversarial Network (SRGAN) can effectively restore the texture details of the image. However, experimentally verified that the texture details of the image recovered by the SRGAN are not robust. In order to get super-resolution reconstructed images with richer high-frequency details, we improve the network structure and propose a super-resolution reconstruction algorithm combining wavelet transform and Generative Adversarial Network. The proposed algorithm can efficiently reconstruct high-resolution images with rich global information and local texture details. We have trained our model by PyTorch framework and VOC2012 dataset, and tested it by Set5, Set14, BSD100 and Urban100 test datasets.

## Keywords
Wavelet transform, Generative Adversarial Network, Deep learning,

Super-resolution reconstruction

## 1. INTRODUCTION
Image super-resolution reconstruction is a digital image processing technique that reconstructs LR images into HR images [3, 4]. Super-resolution reconstruction technology has broad development prospects as a research hotspot in the field of image processing [5, 6]. In recent years, with the development of deep learning, the super resolution reconstruction algorithms based on deep learning have been significantly improved in reconstructing image clarity [5, 7, 11]. The learning-based super-resolution algorithm needs to learn the mapping relationship between high-resolution and low-resolution, and then realize image super-resolution reconstruction through mapping relationship. The typical methods contain Super-Resolution Convolutional Neural Network (SRCNN)[11] and Super-Resolution Using a Generative Adversarial Network (SRGAN)[8]. GAN is a model that optimizes the network by competing through the generation model and the discriminant model[3]. Using GAN in super-resolution reconstruction can improve the accuracy of the image. The author also proposes a perceptual loss function that considers human visual senses [8], which is one of the important reasons why the reconstructed image of the SRGAN looks clear. However, the objective evaluation results such like MSE and PSNR of reconstructed images of SRGAN are lower than other deep learning based SR algorithms.

The decomposition result of the wavelet transform can reflect the high frequency information of images of different resolutions. The starting point of this study is to improve the exact high frequency information in the LR image, so we fuse the wavelet into the GAN model. Because the wavelet has the characteristics of decomposing the high-frequency details and low-frequency approximations in the image and separately representing them, we can use the network to predict the wavelet coefficients, which is beneficial to restore the texture details of the image and improve the clarity of the reconstructed image.

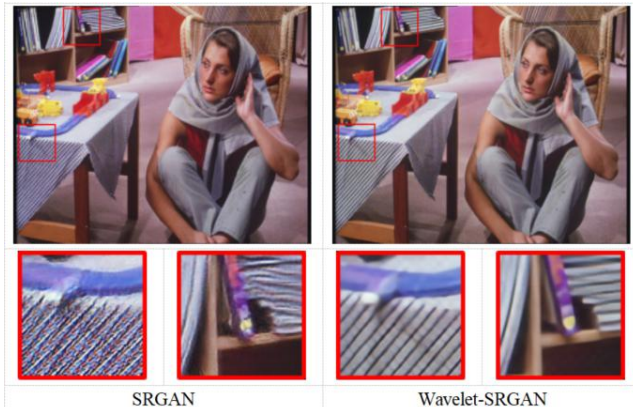

Figure 1: Result of SRGAN and our Wavelet-SRGAN respecttively used on " Barbara". The image that processed by Wavelet-SRGAN with more texture detail.

In order to improve the objective evaluation of the reconstructed image of SRGAN and generate HR images with richer texture detail information, we propose a super-resolution reconstruction algorithm based on wavelet and GAN. In addition, in order to improve the objective evaluation of the SRGAN , we add two evaluation criteria — FSIM and UIQ, in the case of MSE and PSNR. FSIM judges the characteristics of the reconstructed image; UIQ evaluates the image from the linear correlation loss, brightness distortion and contrast distortion of the image.

The main contributions of this paper are the following:1).Propose a first approach to combine the complementarity of information (into low and high frequency sub-bands) in the wavelet domain with GAN. Specifically, wavelet promote sparsity and structural information of the image. 2) We compare the feature distribution of the original HR image to determine the texture feature restoration of the reconstructed image in the case of PSNR and SSIM [14] do not accurately determine the texture restoration.

## 2. RELATED WORK
The deep learning reconstruction algorithm based on deep learning is a kind of method based on learning algorithm with high reconstruction quality and fast reconstruction speed. In 2016, Kim et al. [5] proposed a Deeply-Recursive Convolutional Network (DRCN) based on recurrent neural network, and solved the problem of increased parameters caused by the increase of network depth by using the skip connection structure. In [6], the author proposes an Efficient Sub-Pixel Convolutional Neural Network (ESPCN) based on sub-pixel convolution structure, which effectively improves the feature of extracting features directly on LR images. The operating efficiency of the network. In general, the loss function in the depth learning based algorithm only considers the pixel distortion of the image, and often the algorithm cannot effectively restore the texture details of the image because the visual quality of the image is neglected.

In 2014, Goodfellow et al. [7] first proposed the concept of GAN, including the generation model and the discriminant model in the GAN network framework. GAN is unstable during the training process and has limitations in some application areas. In 2017, the SRGAN algorithm proposed by Ledig et al. [8] used the improved GAN model and the perceptual loss function to recover the texture details of the image.

Akbarzadeh et al. [9] constructed a bi-cubic interpolation and stationary wavelet transform algorithm, which obtained better reconstruction results than the interpolation algorithm. Guo et al. [10] proposed that the wavelet coefficients and wavelet residuals are used as input and output of the network, simplifying the mapping relationship that the network needs to learn. Later, Huang et al. [12] proposed a super-resolution reconstruction algorithm that combines wavelet transform with convolutional neural networks. In this algorithm, the network model realizes the reconstruction of HR images by learning the mapping relationship between wavelet decomposition coefficients. Applying the wavelet transform to the neural network helps to restore the texture details of the image, thereby reconstructing a higher quality super-resolution reconstructed image.

## 3. Proposed method

In order to enable GAN to reconstruct more accurate image texture details, we propose a super-resolution reconstruction algorithm that combines wavelet and GAN. Make use of the advantage of GAN that reconstruct the texture details of images and enhance image global information consistency by training high frequency and low frequency components of wavelet decomposition in the network and improve the recovery of local details of images.

## 3.1 The framework of fusion of Deep learning mechanism and wavelet

Goodfellow first proposed the concept of Generative Adversarial Networks[7].GAN is based on a game theory. When the network is training, the competition between the generative Model and the discriminative model enables the network to learn deeper feature information and improve the training efficiency of the network. A large number of applications in the field of image style conversion migration, text conversion images, super-resolution reconstruction, and other digital image processing fields[1, 2].When using GAN for super-resolution reconstruction, the generative network in the model is used to process the LR image to generate an HR image, and the Adversarial network is used to determine whether the image is the original HR image. In the network structure, when the image generated by the generative network can "cheat" the Adversarial network and discriminate it as the original image, the output image is the result of the GAN model.

Wavelet has the characteristics of extracting useful information from signals, and multi-resolution analysis of signals through various arithmetic methods. It has been proved that wavelet is an efficient multi-resolution image representation tool [15-18], and its decomposition results can reflect high-frequency and low-frequency information of images with different resolutions. The image is decomposed into independent information by wavelet Transformation. We can regard the wavelet coefficient as the eigenvector of the image, and the high-frequency component can reflect the frequency variation in different directions, which means that we can get more texture detail of the image.

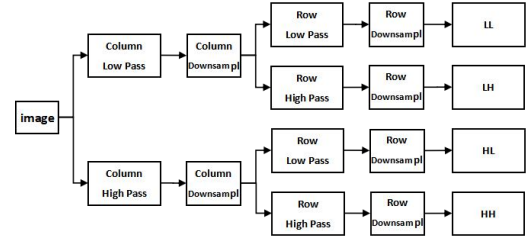

$$Y_{1,L[m,n]} = \sum_{k=0}^{K-1} X[m,2n-k]l[k] \quad \begin{cases} X_{1,LL[m,n]} = \sum_{k=0}^{K-1} Y_{1,L}[2m-k,n]l[k] \\ X_{1,HL[m,n]} = \sum_{k=0}^{K-1} Y_{1,L}[2m-k,n]h[k] \\ X_{1,LH[m,n]} = \sum_{k=0}^{K-1} Y_{1,H}[2m-k,n]l[k] \\ X_{1,HH[m,n]} = \sum_{k=0}^{K-1} Y_{1,H}[2m-k,n]h[k] \end{cases}$$

$$Y_{1,H[m,n]} = \sum_{k=0}^{K-1} X[m,2n-k]h[k]$$

（1）

Wavelet packet Transformation can decompose an image into a series of wavelet decomposition coefficients with the same size. One digital image can be viewed as a 2D signal with index [m,n], where x[m,n] is the pixel value located at mth column and nth row. The 2D signal x[m,n] can be regarded as 1D signals among the rows x[m, :] at a given mth column and among the columns x[:, n] at a given nth row. If the input LR image is X [m, n], we set the high-pass and low-pass filters in the wavelet packet Transformation to h[k] and l[k], respectively(As shown in formula 1 ).

Figure 2: The procedure of fast transform. FWT use low-pass and high-pass decomposition fillers iteratively to compute wavelet coefficients.

When performing 1D wavelet packet Transformation , it is necessary to perform high-pass and low-pass filtering and down-sampling processing on the nth, then perform high-pass and low-pass filtering and down-sampling processing in the mth.

We overlay a residual network with the same structure in the generative network. Residual network, is the most important part of the generative model, can handle the deep network effectively and reduce the training difficulty of GAN. In the original residual network, there are two convolution layers and two normalization layers and the activation function *Relu*. Inspired by Bee Lim[13],we change the structure of the residual network. Specifically, we remove the normalization layer in the residual network, because the normalization layer consumes a large amount of memory when training the network. It's a effective way to improve the efficiency and performance when training network. In addition, we also replace the original Relu activation function with ParametricReLU.

We train the wavelet coefficients of images to reach the goal of super-resolution image reconstruction. Discrete wavelet packet decomposetion and reconstruction of two-dimensional images often use Haar wavelet as wavelet basis function. Haar wavelet is the earliest orthogonal wavelet basis function with tight support. Its function is simple and the application is the most extensive, especially its better symmetry. It can effectively help us avoid phase distortion during image decomposition.

## 3.2 Loss function

As we all know, In deep learning, the training is a process of continuously reducing the loss value through iteration until the optimal solution is obtained. In general, MSE is commonly used as a loss function in super-resolution reconstruction algorithms because the value of PSNR is related to MSE in the objective evaluation criteria for image quality. Therefore, we can get higher PSNR when using MSE as a loss function. However, the edges of the image reconstructed by this loss function with obvious sawtooth and the image is visually blurred. Ledig[8] thought that the loss function should be set in consideration of human perception, and propose a perceptual loss function defined as the sum of content loss $l_C$ and anti-loss $l_A$, which can be expressed as equation (1).

$$loss = l_C + l_A \quad (1)$$

In this paper, we added the wavelet coefficients to predict the loss based on the perceptual loss function proposed by Ledig, which can be expressed as formula (2).

$$loss = l_C + l_A + l_{wavelet} \quad (2)$$

The content loss lC represents the mean-square error(MSE) between the high-level features of the image extracted through the 19-layer VGG network, which shown in formula (4-3). By using the error function between the high-level features of the image as the loss function of the network, the texture information in the reconstructed image can be effectively improved, and the image can be seen more clearly.

$$l_{C/i,j} = \frac{1}{W_{i,j}H_{i,j}} \sum_{x=1}^{W_{i,j}} \sum_{y=1}^{H_{i,j}} (\phi_{i,j}(Y)_{x,y} - \phi_{i,j}(G(X))_{x,y})^2 \quad (3)$$

In the formula, $\varphi_{i,j}$ represents the feature map obtained by convolution of the j-th layer after the maximum pooling of the i-th layer in the VGG19 network. In our paper, the loss of wavelet coefficient prediction wavelet can be expressed as the formula (4).

$$l_{wavelet} = \sum_{i}^{N_w} \alpha_i (w_i^{HR} - w_i^{SR})^2 \quad (4)$$

$N_w$ is the number of wavelet coefficients decomposed by the image by wavelet packet transform, $w_i^{HR}$, $w_i^{SR}$ are the wavelet coefficients of the HR image and the reconstructed image, respect-ively.

$$l_A = \sum_{n=1}^{N} -\log D(G(X)) \quad (5)$$

The adversarial loss lA is based on the probability of identifying the output of the network, and is generally expressed as equation (5).

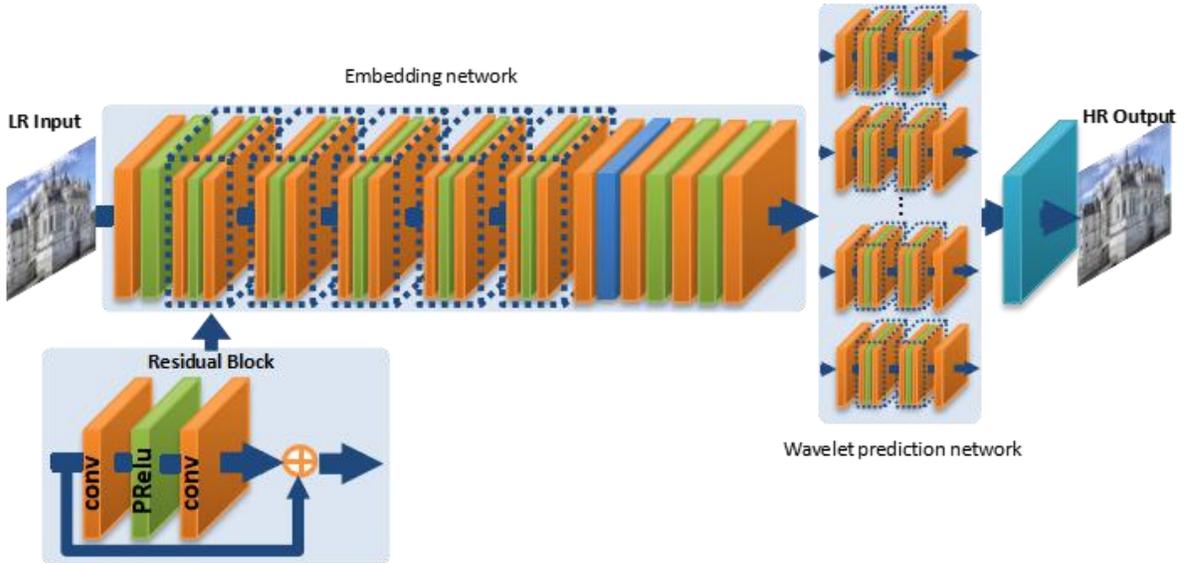

Figure 3. The architecture of our wavelet-based super-resolution Generative Adversarial Networks .The Residual Block we use ParametricReLU. All the convolution layers have the same filter map size of 3x3 and each number below them defines their individual channel size. Skip connections exist between every two convolution layers (except the first layer) in the embedding and wavelet predicting nets.

## 3.3 Algorithm

The training process of the GAN model incorporating wavelets also needs to be alternately optimized between the generated model and the discriminant model. We only need to add the loss of wavelet coefficient prediction to the loss function of the generated model during the optimization process. The complete fusion wavelet and GAN model super-resolution reconstruction algori-thm is as follows.

Table 1: The wavelet-based SRGAN algorithm

```
Input: Learning rate l, Batch size a, Epoch N, Generative model param g,
Adversarial model param d
Output: new generative model param g', adversarial model param d'
From epoch 1→N
    Select a-factor HR images from the training set randomly
    Downsample a-factor HR images to obtain corresponding LR images
    Calculate the gradient→The loss of adversarial network l_d
    Update the param d' of adversarial network
    2dDWT for the HR images→a×3×16 HR params
    2dDWT for the LR images→a×3×16 LR params
    Calculate the gradient→the loss of Generative network l_c+l_wavelet
    Update the param g' of generative network
End epoch
```

## 4. EXPERIMENT

### 4.1 Dataset

In our experiment, we use the VOC2012 dataset as the training set for our model. The VOC2012 dataset is an image dataset for super-resolution reconstruction which includes 16,700 training images and 425 test images. VOC2012 dataset includes a total of 20 sub-categories in the four categories "Person", "Animal", "Vehicle", and "Indoor", which are rich in variety and suitable for super-resolution reconstruction. We tested the trained network using Set5 dataset, Set14 dataset, BSD100 dataset, and Urban100 dataset.

### 4.2 Training Settings

In training process, the LR image is obtained by performing bicubic interpolation(The downsampling factor is 4) on 16700 HR images. During training, the HR image is randomly clipped into a sub-image of size 88×88. The network reconstructs the image by learning the mapping relationship between the wavelet coefficients of the LR image and the wavelet coefficients of the HR image. The related network parameters are set as follows: the size of the batch size in the network is set to 16, the network learning rate is 0.0002 and the number of iterations is 313,200.

### 4.3 Exparimental results and analysis

We compare the reconstruction performance of this research method with the SRGAN algorithm. We reproduced the SRGAN algorithm on the VOC2012 training set. Due to the different training sets, the test results of the SRGAN algorithm are slightly different from the test results in the author's original text. In order to accurately evaluate the quality of reconstructed images. In our experiment, not only PSNR, SSIM, FSIM and UIQ are used as evaluation indexes, but also LBP is used to extract image features to judge the texture restoration degree of the reconstructed image.

Table 2 shows the PSNR and SSIM test results for the two algorithm reconstruction results. The data in the table is the average of the results of the test of the Set5 data set, the Set14 data set, the BSD100 data set, and the Urban100 data set downsampled by double-quadratic interpolation 4 times. According to the table 4, our algorithm has the highest PSNR and SSIM evaluation for each data set reconstruction result. The minimum average of the PSNR and SSIM test results is 0.18 dB and 0.018 higher than the SRGAN.

Table 2: Test results of PSNR and SSIM

| Dataset | SRGAN (PSNR, SSIM) | Wavelet-SRGAN (PSNR, SSIM) |
|---|---|---|
| Set5 | 29.27, 0.852 | 30.63, 0.897 |
| Set14 | 25.95, 0.740 | 27.04, 0.770 |
| BSD100 | 25.25, 0.710 | 26.91, 0.728 |
| Urban100 | 23.94, 0.718 | 24.12, 0.743 |

Table 3 shows the test results of FSIM and UIQ for the pictures obtained by the two reconstruction algorithms. To evaluate the quality of the reconstructed image more precisely, we use FSIM and UIQ to test reconstructed images and analyze the results. The closer the values of FSIM and UIQ are to 1, the more similar the restored image is to the original image. Similarly, the data in the table is the average. According to the table, both the FSIM and UIQ of Wavele-SRGAN are better than SRGAN. Further more, the average value of FSIM is 0.06 higher than the SRGAN, and the average value of the UIQ is the same as the SRGAN.

Table 3: Test results of FSIM and UIQ

| Dataset | SRGAN (FSIM, UIQ) | Wavelet-SRGAN (FSIM, UIQ) |
|---|---|---|
| Set5 | 0.906, 0.935 | 0.929, 0.935 |
| Set14 | 0.891, 0.973 | 0.899, 0.978 |
| BSD100 | 0.826, 0.963 | 0.832, 0.984 |
| Urban100 | 0.949, 0.967 | 0.957, 0.967 |

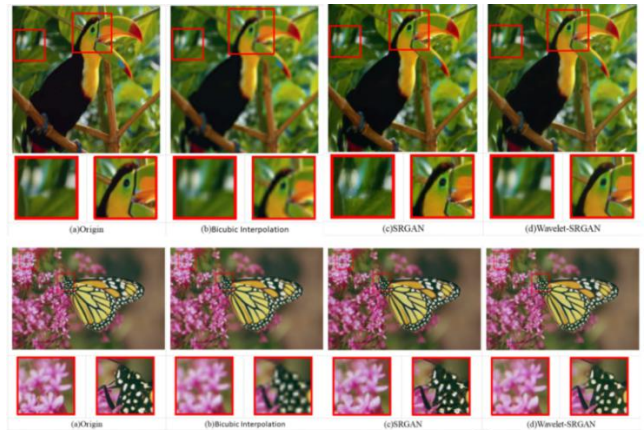

Figure 4. Super-resolution reconstruction of "bird" and "monarch" by different algorithms

Figure 4 shows the results of super-resolution reconstruction with 4x magnification of the images "bird" and "monarch" by different algorithms and their partially enlarged images. In light of the figure, we can find that the bicubic interpolation can't fix the high frequency information in the image well. And the output image's edge is jagged, the visual blur is more obvious. The reconstructed images of SRGAN are visually clear because of its ability of reconstructing the details effectively. However, there are artifacts in the reconstructed images after partial enlargement, especially the "barbara". It's obvious that the visual quality of the reconstructed image in Wavelet-SRGAN is the best, due to wavelet packet transform restores texture details at the edge of the image. The partially enlarged image is also the sharpest and closest to the original image.

## 5. CONCLUSIONS

The GAN-based model can reconstruct HR images with clear textures. Because of the ability of wavelet packet transform that decomposing high-frequency and low-frequency details of a image and representing them separately. In order to improve the quality of reconstructed images, we propose super-resolution reconstruction algorithm that combines wavelet transform and GAN. The individual training of the wavelet decomposition coefficients of the image through the network makes the texture of the reconstructed image more natural.

The experimental results show that the proposed method can effectively improve the shortcomings of the objective evaluation of the reconstructed image of the SRGAN algorithm, and at the same time generate reconstructed images with more realistic texture details. This shows that our proposed algorithm takes advantage of wavelet packet transform and GAN in the field of super-resolution reconstruction.